\documentclass[prd,twocolumn,aps,floats, %superscriptaddress,longbibliography,
floatfix,showpacs,showkeys,amsmath,amssymb,nofootinbib]{revtex4-1}

\usepackage{dcolumn}% Align table columns on decimal point
\usepackage{bm}% bold math
\usepackage{txfonts}% bold math
\usepackage{pxfonts}% bold math
\usepackage{graphicx}
\usepackage{bigints}
\usepackage{psfrag}
\usepackage{times}

\begin{document}
\title{Acoustic Hawking radiation as a tunnelling effect in 
Michel accretion}

\author{Nisha Jangid}\email{202121002@dau.ac.in}
%\affiliation{School of Physical Sciences, Indian Association
%for the Cultivation of Science, Jadavpur, Kolkata 700032, India} 
%\altaffiliation{Harish Chandra Research Institute, Chhatnag 
%Road, Jhunsi, Allahabad 211019, India}

%\author{Arnab K. Ray}\email{arnab\_kumar@daiict.ac.in}
\author{Arnab K. Ray}\email{arnab\_kumar@dau.ac.in}
\affiliation{Dhirubhai Ambani University, School of Technology, 
Gandhinagar 382007, Gujarat, India}
%\affiliation{Department of Physics, Jaypee University of 
%Engineering and Technology, %A-B Road, 
%Raghogarh, Guna 473226, Madhya Pradesh, India}

\date{\today}

\begin{abstract}
Michel accretion becomes transonic at the saddle point of
a dynamical system. An %time-dependent 
Eulerian perturbation on the steady inflow 
produces the metric of an acoustic black hole. As a
high-frequency travelling wave the perturbation 
does not destabilize the steady inflow. 
% but the geometry of Schwarzschild spacetime affects the wave. 
Acoustic waves propagating outwards 
against the fluid inflow are blocked at the sonic barrier 
but can tunnel through it 
%horizon of the acoustic black hole 
with an exponentially decaying
amplitude. The Hawking temperature and the frequency of the 
Hawking phonons are enhanced by the spacetime geometry.  
\end{abstract}

\pacs{98.62.Mw, 97.60.Lf, 04.70.Dy} 

\keywords{Infall, accretion; Black holes; 
Quantum aspects of black holes} 

\maketitle

\section{Introduction} 
\label{sec1}
Spherically symmetric accretion is a textbook model of a 
transonic flow in astrophysical fluid dynamics~\citep{shap83,fkr02}. 
Originally introduced in the Newtonian framework of space and time 
by~\citet{bon52}, the model was later advanced to the general
relativistic Schwarzschild spacetime by~\citet{mich72}. 
The common physical system in both models is a non-self-gravitating
compressible fluid that undergoes spherically symmetric transonic
infall, driven by the gravity of a centrally-located accretor 
(stars, neutron stars, black holes, etc.). The transonic nature
of the flow in both cases is a settled fact~\citep{bon52,mich72,
beg78,mon80,ds01,rb02,mrd07,rr07,sr14}. Such flows originate far  
away from the accretor, where the flow velocity has a 
vanishingly small subsonic value but becomes highly  
supersonic as it nears the accretor. Thus, with the global 
flow regime 
being subsonic at the outer boundary and supersonic at the
inner boundary, the flow must become transonic 
at an intermediate radius~\citep{bon52,rb02}. This radius 
defines the surface of a spherical sonic barrier.
Acoustic waves in the supersonic region are 
completely trapped within this spherical sonic surface.
As a result the region
bounded by the surface becomes a spherical
acoustic black 
hole~\citep{mon80,wgu81,wgu95,mvis98,vis98,blv11},
and the sonic surface 
becomes a sonic horizon -- the acoustic analogue of the event
horizon of a general relativistic black hole.

Quantum mechanical effects make it possible for general 
relativistic black holes to emit thermal radiation~\citep{swh74,swh75}. 
This radiation -- Hawking radiation -- is a kinematic effect, 
generic to Lorentzian geometries with an event horizon~\citep{mvis98}.  
Therefore, fluid systems with an analogue event horizon can   
emit Hawking quanta, the feasibility 
of which is owed to a closeness
between the behaviour of fields near black holes and waves in
transcritical fluid flows~\citep{wgu81,vis98}.
Thus it happens that analogues of gravity in a diverse range 
of physical systems %(not necessarily only fluidic) 
are now well established (see~\citep{blv11} for a 
review), astrophysical accretion being one of 
them~\citep{mon80,bil99,das04,dbd05,abd06,dbd06,akr20}.
The basic physical characteristics of~\citet{bon52} accretion 
and~\citet{mich72} accretion, namely, spherically symmetric, 
compressible, conserved, convergent and irrotational, suit 
them both 
as three-dimensional transonic potential flows. These 
aspects of spherically symmetric accretion make 
it a subject of interest from the perspective of analogue 
gravity as well~\citep{mon80,bil99,das04,dbd05,akr20}, especially 
since natural examples of such flows are quite rare. 

Tunnelling through the event horizon is one means by which 
Hawking radiation may be realized. It is understood that while
absorption into black holes occur classically, thermal emission 
from black holes is a quantum phenomenon~\citep{cgm08}, 
like tunnelling through a barrier. 
Studies on tunnelling have been reported  
in general relativity (see~\citep{pw2k,pm07,cgm08} and 
references therein) as well as in fluid 
analogues of gravity~\citep{vol06,jkb17,akr20,njar25}. 
Specific to astrophysical accretion, it is known that 
if viscosity is introduced as a perturbative effect about 
the~\citet{bon52} solution, then the Lorentzian structure 
of the acoustic metric is lost, but then again, the same viscous
dissipation
facilitates the tunnelling process through the acoustic
horizon by softening it~\citep{akr20}. It is also known 
that in~\citet{mich72} accretion the coupling of the geometry
of Schwarzschild spacetime with the flow acts like a dissipative
effect and causes the Lorentzian symmetry to be 
broken~\citep{ncbr07}. 

In the present work, our objective is to study the effect of 
the aforementioned coupling on the tunnelling of acoustic waves 
through the sonic horizon of an acoustic black hole in~\citet{mich72} 
accretion. 
%The layout of this paper is as follows. 
In Sec.~\ref{sec2} we set 
down the relevant fluid equations in Schwarzschild geometry, 
with the necessary variables expressed explicitly as functions 
of time and the radial coordinate. 
In Sec.~\ref{sec3} we discuss the steady conditions 
related to~\citet{mich72} accretion. 
We show that the fixed point of the~\citet{mich72} 
inflow is a saddle point of a dynamical system, a result 
that generalizes a similar result~\citep{rb02} 
related to~\citet{bon52} accretion.   
In Sec.~\ref{sec4} we use a linearized Eulerian perturbation 
about the steady background flow 
to derive a wave equation and establish the metric of an 
acoustic black hole at the transonic point. We 
show that the coupling of the spacetime curvature and
the perturbed field breaks the symmetry of the acoustic metric. 
A dispersion relation, derived from the wave equation, 
shows that spacetime curvature subdues 
the group velocity and can cause 
%affects the amplitude such in a way that 
%derived from the wave equation 
%shows how the amplitude of the wave and its group 
%velocity are affected by the spacetime curvature. 
a possible divergence of the amplitude. % is also indicated here. 
In Sec.~\ref{sec5} we carry out a {\it WKB} analysis to show 
that high-frequency acoustic waves with a slowly-varying 
amplitude do not destabilize the steady flow. 
In Sec.~\ref{sec6}, we apply Cauchy's residue analysis 
about the singularity at the sonic horizon and show how 
acoustic waves that are blocked at the sonic barrier can tunnel 
through it with 
a decaying amplitude. Spacetime curvature enhances the analogue
Hawking temperature and the frequency of the emitted Hawking phonons.  

%\newpage
\section{The equations of the flow} 
\label{sec2} 
The time-dependent equations for~\citet{bon52} accretion
are framed in Newtonian space and time~\citep{rb02}.  
Equivalent equations for~\citet{mich72} accretion 
derive from a conserved spherically 
symmetric non-self-gravitating gas inflow in Schwarzschild 
geometry. The line element for this geometry~\citep{dem85},
with the unit of $c=1$ and with the radial distance scaled by $2GM$, 
is
\begin{equation}
\label{line}
{\mathrm d}s^2 =f(r){\mathrm d}t^2 - \frac{{\mathrm d}r^2}{f(r)} 
-r^2 \left({\mathrm d}\theta^2 +\sin^2\theta \,{\mathrm d}\phi^2\right),  
\end{equation}
in which the metric coefficient for a black hole of mass, $M$, is 
\begin{equation}
\label{statmetric}
f(r) = 1-\frac{2GM}{r}. 
\end{equation}
While~\citet{bon52} accretion is driven 
mechanically by 
the gravity of a centrally-located mass, in~\citet{mich72} accretion
the gravitational effect is due to the curved %spacetime of 
Schwarzschild geometry. This geometry will be static  
if $M$ is constant, since it will make 
the metric coefficients in Eq.~(\ref{line}) static as well.
%, then Eq.~(\ref{line}) ensures a static geometry. 

The continuity condition for a spherically symmetric inflow of a 
compressible fluid in the geometry implied by Eq.~(\ref{line}) is  
\begin{equation}
\label{cont}
\left(\rho v^{\mu}\right)_{;\mu}=0, 
\end{equation}
where $\rho$ is the particle number density of the fluid
and $v^{\mu}$ is the fluid four-velocity, which observes
$v^{\mu}v_{\mu} = 1$. The radial flow velocity, $v^1 \equiv u(r,t)$, 
and the local density, $\rho (r,t)$, can be related to each other
through Eqs.~(\ref{line}) and~(\ref{cont}). This leads to 
the continuity equation~\citep{ncbr07},  
\begin{equation} 
\label{modcon}
\frac{\partial}{\partial t}\left(\frac{\rho \sqrt{f+u^2}}{f}\right) 
+ \frac{1}{r^2}\frac{\partial}{\partial r}\left(\rho ur^2\right) =0.  
\end{equation}

To solve for $u(r,t)$ and $\rho(r,t)$ we need another mathematical 
relation between them. Such a relation can be derived from the 
energy-momentum tensor of an ideal fluid~\cite{dem85}, 
\begin{equation}
\label{stressenergy}
T^{\mu\nu} =\left(\epsilon + p\right)v^{\mu}v^{\nu}-pg^{\mu \nu},  
\end{equation}
in which $p$ is the pressure and $\epsilon$ is the energy density. 
Both are thermodynamically related to each other through 
\begin{equation}
\label{thermo}
\mathrm{d}\left(\frac{\epsilon}{\rho}\right) 
+ p\mathrm{d}\left(\frac{1}{\rho}\right) = T{\mathrm d}S,
%\frac{{\mrm d}\epsilon}{{\mrm d}\rho} = \frac{\epsilon + p}
%{\rho} + \rho T \frac{{\mrm d} {\mrm S}}{{\mrm d}\rho}
\end{equation}
where $T$ is the temperature and $S$ is the specific entropy.
In the isentropic limit, the speed of
sound, $a$, is defined by
\begin{equation}
\label{sound}
a^2 = \frac{\partial p}{\partial \epsilon}\bigg{\vert}_{S} . 
\end{equation}

Now we apply the condition of energy-momentum 
conservation, ${T^{\mu \nu}}_{; \nu} =0$, on Eq.~(\ref{stressenergy}). 
Thereafter, making use of Eqs.~(\ref{thermo})~and~(\ref{sound}) for 
the isentropic case, %($\mathrm{s}$ is constant), 
along with
the line element in Eq.~(\ref{line}), we get~\citep{ncbr07} 
\begin{multline}
\label{modenermom}
\frac{\sqrt{f+u^2}}{f} \frac{\partial u}{\partial t} +
u\frac{\partial u}{\partial r} +\frac{1}{2} 
\frac{\partial f}{\partial r} \\
%- \frac{v \sqrt{f + v^2}}{f^2} \frac{\prt f}{\prt t} 
+ \frac{a^2}{\rho} \left[\frac{u \sqrt{f+u^2}}{f} 
\frac{\partial \rho}{\partial t}
+\left(f+ u^2\right) \frac{\partial \rho}{\partial r}\right]=0.   
\end{multline}

The pressure, $p$, depends on the density, $\rho$, through
a polytropic equation of state, $p = k\rho^{\gamma}$, where $k$
and $\gamma$ are constants, the latter being the polytropic
exponent~\citep{chan39}. With the polytropic equation of state, 
we can show from Eqs.~(\ref{thermo}) and~(\ref{sound}) that when 
the fluid is isentropic, $a$ depends on $\rho$ as  
\begin{equation}
%\label{conden}
\label{soundgen} 
a^2 = \frac{\gamma k\rho^{\gamma -1}}{{\bar{\mu}} 
+ n\gamma k\rho^{\gamma -1}}
%\rho =\left[\frac{a^2}{\gamma k \left(1-na^2\right)}\right]^n 
\end{equation}
in which $n=(\gamma -1)^{-1}$ is the polytropic index~\cite{chan39}
and $\bar{\mu}$ is the mean baryon mass~\citep{shap83}.
With $a(\rho)$ defined in Eq.~(\ref{soundgen}), we
see now that Eq.~(\ref{modenermom}) can be expressed
entirely in terms of $u(r,t)$ and $\rho (r,t)$. 
Thus, Eqs.~(\ref{modcon}),~(\ref{modenermom}) and~(\ref{soundgen})
together give a complete mathematical description of a spherically 
symmetric non-self-gravitating gas inflow 
%that does not change with time.  description of the spherically
%symmetric non-self-gravitating gas inflow 
in Schwarzschild geometry. %, set up by a black hole of constant mass. 

%\newpage
\section{Steady transonic accretion}
\label{sec3}
In the steady state the radial flow becomes free of explicit 
time-dependence, whereby we set $\partial/\partial t \equiv 0$. 
This condition allows us to integrate the spatial part of
Eq.~(\ref{modcon}) over the full spherical front of the inflow 
to obtain 
\begin{equation} 
\label{intecon}
4\pi {\bar{\mu}}\rho ur^2 =-{\dot{m}}. 
\end{equation} 
On the right hand side of the foregoing equation, the constant
of the motion, $\dot{m}$, is the steady accretion 
rate and the negative sign with $\dot{m}$ implies an inward 
flux of matter due to $u(r) <0$.  
As for Eq.~(\ref{modenermom}), in the steady limit it appears 
as %~\citep{shap83}
\begin{equation} 
\label{ordmom} 
u\frac{{\mathrm d}u}{{\mathrm d}r} 
+\frac{1}{2} \frac{{\mathrm d}f}{{\mathrm d}r}
+\frac{a^2}{\rho} \left(f+u^2\right) 
\frac{{\mathrm d}\rho}{{\mathrm d}r} =0, 
\end{equation} 
whose integral solution can be found by eliminating the spatial 
derivative of $\rho$ with the help of Eq.~(\ref{soundgen}). This
will lead to %~\citep{shap83}  
\begin{equation} 
\label{intenermom} 
\left(1-\frac{2GM}{r} +u^2\right)
\left(1+\frac{a^2}{\gamma -1 -a^2}\right)^2 = 
\left(1+\frac{a_\infty^2}{\gamma -1 -a_\infty^2}\right)^2, 
\end{equation} 
with the constant on the right hand side having been fixed  
with the boundary conditions, $u(\infty) \longrightarrow 0$
and $a(\infty) \longrightarrow a_\infty$, the ambient 
value of the speed of sound.  

Steady spherically symmetric accretion admits a monotonic 
and continuous  inflow solution that originates
with a subsonic velocity far away from the accretor and reaches 
the accretor with a supersonic velocity. This is known for
both~\citet{bon52} accretion and~\citet{mich72} accretion~\citep{beg78}. 
Obviously then, the solution must become transonic at some 
critical radius. The critical conditions 
can be known by recasting Eq.~(\ref{ordmom}) as 
\begin{equation} 
\label{dudr} 
\frac{{\mathrm d}u^2}{{\mathrm d}r} = 
\frac{2u^2 \left[2a^2(f+u^2)-(GM/r)\right]}
{r\left[u^2 -a^2\left(f+u^2\right)\right]}. 
\end{equation} 
When the flow becomes transonic, both the numerator and the 
denominator on the right hand side of Eq.~(\ref{dudr}) vanish 
simultaneously. The resulting critical flow coordinates, 
labelled with the subscript ``c", are 
\begin{equation} 
\label{critcon1}
u_{\mathrm c}^2=a_{\mathrm c}^2 \left[f(r_{\mathrm c})
+u_{\mathrm c}^2\right] = \frac{GM}{2r_{\mathrm c}}
= \frac{a_{\mathrm c}^2}{1+3a_{\mathrm c}^2}.  
%\begin{equation} 
%\label{critcon2}
%u_{\mathrm c}^2=
%\frac{a_{\mathrm c}^2}{1+a_{\mathrm c}^2} 
%= \frac{GM}{2r_{\mathrm c}}. 
\end{equation}
The last term in the Eq.~(\ref{critcon1}) explicitly renders 
$u_{\mathrm c} \equiv u_{\mathrm c} (a_{\mathrm c})$ and 
$r_{\mathrm c} \equiv r_{\mathrm c} (a_{\mathrm c})$.  
Inserting these two conditions in Eq.~(\ref{intenermom}) will 
fix $a_{\mathrm c}$ in terms of the fixed parameters, $\gamma$ 
and $a_\infty$. This will give
\begin{equation} 
\label{fixing} 
\left(1+3a_{\mathrm c}^2\right) 
\left(\gamma -1-a_{\mathrm c}^2\right)^2 = 
\left(\gamma -1-a_{\infty}^2\right)^2. 
\end{equation} 
With $a_{\mathrm c}$ thus fixed, both $u_{\mathrm c}$ and 
$r_{\mathrm c}$ in Eq.~(\ref{critcon1}) also become fixed coordinates
of the flow. 
We note that Eq.~(\ref{fixing}) is a cubic equation in 
$a_{\mathrm c}^2$. As such $a_{\mathrm c}^2$ will have three roots, 
of which at least one must be real and positive, 
corresponding to a single physical
transonic point. We note that for $\gamma > 5/3$ multiple critical 
points can occur~\citep{beg78,mrd07,cms16}, 
but not all of them will be physically meaningful. However, here we
avoid such outcomes by restricting $\gamma$ within the range of 
$1 < \gamma < 5/3$, the lower limit corresponding to the isothermal 
case and the 
upper limit to the adiabatic case, respectively~\citep{chan39}. 

To understand the nature of the critical point, now we follow an
established method in accretion 
studies~\citep{rb02,ap03,crd06,rbcqg07,mrd07,gkrd07,rr07,rnr09,
bbdr09,nard12,cms16} and decompose Eq.~(\ref{dudr}) as a coupled 
first-order autonomous dynamical system~\citep{stro}. 
The resulting coupled set is 
\begin{align} 
\label{dyna1} 
\frac{{\mathrm d}u^2}{{\mathrm d}\tau}& = 
2u^2 \left[2a^2(f+u^2)-\frac{GM}{r}\right] \nonumber \\
\frac{{\mathrm d}r}{{\mathrm d}\tau}& = 
r\left[u^2 -a^2\left(f+u^2\right)\right], 
\end{align} 
in which $\tau$ is a mathematical parameter. The expressions on 
the right hand sides of Eqs.~(\ref{dyna1}) vanish at the fixed 
point. We perturb about the fixed point coordinates according to
the scheme, $u^2 = u_{\mathrm c}^2 + \delta u^2$, 
$a^2 = a_{\mathrm c}^2 + \delta a^2$ and 
$r = r_{\mathrm c} + \delta r$. Then using Eqs.~(\ref{soundgen}) 
and~(\ref{intecon}), we express $\delta a^2$ in terms of 
$\delta u^2$ and  $\delta r$. Following this, 
linearizing Eqs.~(\ref{dyna1}) in the perturbed quantities will 
give us a set of coupled autonomous linear equations, 
\begin{align} 
\label{dyna2} 
\frac{{\mathrm d}}{{\mathrm d}\tau} (\delta u^2)& = 
A \delta u^2 + B \delta r \nonumber \\
\frac{{\mathrm d}}{{\mathrm d}\tau} \left(\delta r\right)& = 
C \delta u^2 + D \delta r,  
\end{align}
in which $A$, $B$, $C$ and $D$ are constant coefficients, 
reading as 
\begin{align} 
\label{dyncoeff} 
A& =-D=2u_{\mathrm c}^2 
\left(1-\gamma +3a_{\mathrm c}^2\right) \nonumber \\
B& = \frac{4u_{\mathrm c}^4}{r_{\mathrm c}}
\left(3 -2\gamma + 6a_{\mathrm c}^2\right) \nonumber \\
C& = \frac{r_{\mathrm c}}{2}\left(1+ \gamma -3a_{\mathrm c}^2\right).  
\end{align}
%with the values of $r_{\mathrm c}$, $u_{\mathrm c}^2$ and 
%$a_{\mathrm c}^2$ given by Eqs.~(\ref{critcon1}) and~(\ref{fixing}). 

The stability of the linear system in Eqs.~(\ref{dyna2}) can be
determined from its Jacobian matrix~\citep{stro}. For solutions 
going as $\delta u^2 \sim \exp(\Lambda \tau)$ 
and $\delta r \sim \exp(\Lambda \tau)$, the eigenvalues, $\Lambda$, 
of the Jacobian matrix are  
$\Lambda = ({\mathcal{T}}\pm \sqrt{{\mathcal T}^2 
-4{\mathcal D}})/2$, with the trace, ${\mathcal T} =A+D$, and the 
determinant, ${\mathcal D} = AD -BC$. From Eqs.~(\ref{dyncoeff})
we see that ${\mathcal T}=0$. Solving for the determinant, 
$\mathcal D$, we find the eigenvalues to be  
\begin{equation} 
\label{eigen} 
%\Lambda_{1,2} = \pm 
\Lambda = \pm 
%\frac{a_{\mathrm c}^2}{\left(1+3a_{\mathrm c}^2\right)}
\frac{a_{\mathrm c}^2}{1+3a_{\mathrm c}^2}
\sqrt{2\left(5-3\gamma +9a_{\mathrm c}^2\right)}.  
\end{equation}
With $a_{\mathrm c}^2$ being a 
real positive root of Eq.~(\ref{fixing}), 
both eigenvalues %in Eq.~(\ref{eigen}) 
will be real with opposite 
signs for $\gamma <[(5/3)+3a_{\mathrm c}^2]$. 
Hence, the fixed point of~\citet{mich72} accretion will 
be a saddle point. For the non-relativistic~\citet{bon52} 
accretion, the eigenvalues in Eq.~(\ref{eigen}) will be
$\Lambda = \pm a_{\mathrm c}^2 
\sqrt{2\left(5-3\gamma \right)}$~\citep{rb02}, indicating a 
saddle point again. On comparing the 
eigenvalues in the two cases of~\citet{mich72} accretion 
and~\citet{bon52} accretion, we conclude that the critical point 
of the former has more robust saddle-like properties than 
the latter. Therefore, transonicity is more pronounced in 
the former case than in the latter, inasmuch as transonicity 
implies the continuous passage of a physical solution through the 
saddle point from the subsonic region to the supersonic region. 
This also agrees with the observation that the mass accretion 
rate is greater under relativistic effects than what it can 
be for the~\citet{bon52} accretion model~\citep{mal99,mrd07}.  

%\newpage
\section{An acoustic black hole}
\label{sec4}
The steady solutions of Eqs.~(\ref{modcon}) and~(\ref{modenermom}) 
are two coupled time-independent fields, %that we write as 
$\rho_0(r)$ and $u_0(r)$. About these
steady solutions we impose small time-dependent radial
perturbations as $\rho (r,t)=\rho_0(r)+\rho^{\prime}(r,t)$
and $u(r,t)=u_0(r)+u^{\prime}(r,t)$, with 
the primed quantities being linear-order perturbations 
about the zero-order state. The smallness of the 
perturbations allows their linearization.
Following an Eulerian 
perturbation scheme due originally to~\citet{pso80}, we define a 
variable, $\psi(r,t)=\rho ur^2$, which, in the steady 
limit of Eq.~(\ref{modcon}), becomes a constant of the motion,
$\psi_0 = \rho_0 u_0r^2$. 
It is associated with the steady accretion rate, as we  
see in Eq.~(\ref{intecon}). The 
linearized fluctuations about $\psi_0$ are given by 
\begin{equation}
\label{psifluc1}
\psi^\prime =\left(u_0 \rho^\prime + \rho_0 u^\prime \right)r^2 . 
\end{equation}
Another condition connecting $u^{\prime}$, $\rho^{\prime}$ and
$\psi^{\prime}$ to one another is derived from Eq.~(\ref{modcon}). 
It is 
\begin{equation}
\label{psifluc2}
\frac{\sqrt{f+u_0^2}}{f} \frac{\partial \rho^\prime}{\partial t} 
+\frac{\rho_0 u_0}{f\sqrt{f+u_0^2}} \frac{\partial u^\prime}{\partial t}
=-\frac{1}{r^2} \frac{\partial \psi^\prime}{\partial r}. 
\end{equation}
From Eqs.~(\ref{psifluc1})~and~(\ref{psifluc2}), we get 
$\rho^\prime$ and $u^\prime$ in terms of $\psi^\prime$ as
\begin{equation}
\label{rhofluc}
\frac{\partial \rho^\prime}{\partial t}= %-\frac{1}{r^2} 
-\frac{u_0}{fr^2} \frac{\partial \psi^\prime}{\partial t} 
-\frac{\sqrt{f+u_0^2}}{r^2} \frac{\partial \psi^\prime}{\partial r} 
\end{equation}
and
\begin{equation}
\label{vfluc}
\frac{\partial u^\prime}{\partial t}= 
%\frac{\sqrt{f + v_0^2}}{\rho_0 r^2}
\frac{\left(f+u_0^2\right)}{f \rho_0 r^2} 
\frac{\partial \psi^{\prime}}{\partial t} 
+\frac{u_0 \sqrt{f+u_0^2}}{\rho_0 r^2}
\frac{\partial \psi^\prime}{\partial r}, 
\end{equation}
respectively. 
Now we look for an independent condition on which we
can apply Eqs.~(\ref{rhofluc})~and~(\ref{vfluc}). This condition
comes from Eq.~(\ref{modenermom}). Taking its second-order time 
derivative and applying linearized perturbations in $\rho$ 
and $u$ will lead to  
\begin{multline}
\label{momfluc}
\frac{\sqrt{f+u_0^2}}{f} \frac{\partial^2 u^\prime}{\partial t^2} 
+\frac{\partial}{\partial r}
\left(u_0 \frac{\partial u^\prime}{\partial t}\right) 
+\frac{u_0 \sqrt{f+u_0^2}}{f} \frac{a_0^2}{\rho_0} 
\frac{\partial^2 \rho^\prime}{\partial t^2} \\
+2u_0 \frac{a_0^2}{\rho_0} \frac{\partial \rho_0}{\partial r} 
\frac{\partial u^\prime}{\partial t} 
+\left(f+u_0^2 \right) \frac{\partial}{\partial r} 
\left(\frac{a_0^2}{\rho_0} 
\frac{\partial \rho^\prime}{\partial t}\right) = 0.  
\end{multline}
Thereafter, using Eqs.~(\ref{rhofluc}),~(\ref{vfluc})~and their 
second-order time derivatives in Eq.~(\ref{momfluc}) 
yields a wave equation, 
\begin{multline} 
\label{psieqnmot}
\frac{\partial}{\partial t} \left(h^{tt}
\frac{\partial \psi^\prime}{\partial t} \right) + 
\frac{\partial}{\partial t} 
\left(h^{tr}\frac{\partial \psi^\prime}{\partial r}\right) +  
\frac{\partial}{\partial r} 
\left(h^{rt}\frac{\partial \psi^\prime}{\partial t} \right)
+\frac{\partial}{\partial r} \left(
h^{rr}\frac{\partial \psi^\prime}{\partial r}\right) \\
= \frac{\mathrm d}{{\mathrm d}r} 
\left[\ln \left(f+u_0^2\right)\right]
\left(h^{rt}\frac{\partial \psi^\prime}{\partial t}
+h^{rr}\frac{\partial \psi^\prime}{\partial r}\right), 
\end{multline}
%in which $h^{\alpha \beta}$ are 
in which $h^{tt}$, $h^{tr}$, $h^{rt}$ and $h^{rr}$ are, respectively, 
\begin{align} 
\label{metelem}
h^{tt}& = \frac{u_0 \sqrt{f+u_0^2}
\left(f+u_0^2 -u_0^2 a_0^2 \right)}{f^2} \nonumber \\ 
h^{tr}& = h^{rt} = \frac{u_0^2 \left(f+ u_0^2\right)
\left(1 -a_0^2)\right)}{f} \nonumber \\ 
h^{rr}& = u_0 \sqrt{f+ u_0^2}\left[u_0^2 -a_0^2\left(f+u_0^2 \right)
\right].  
\end{align} 
In the non-relativistic limit both $u_0^2$ and $a_0^2$ become
negligibly small compared to unity, whereas $f \longrightarrow 1$. 
%In that case $h^{tt}=0$, $h^{tr}=h^{rt}=u_0^2$ and 
%$h^{rr}=u_0(u_0^2 -a_0^2)$, whereupon, 
Hence, the right hand side of Eq.~(\ref{psieqnmot}) will 
vanish, rendering it compactly as   
\begin{equation}
\label{compact} 
\partial_\alpha\left(h^{\alpha \beta} \partial_\beta 
\psi^\prime \right)=0, 
\end{equation}
%a form that preserves Lorentz symmetry. 
with the Greek indices, $\alpha$ and $\beta$, running over 
both $t$ and $r$. All the $h^{\alpha \beta}$ in 
Eqs.~(\ref{metelem}) will then become elements of the matrix,  
\begin{equation}
\label{symmat}
h^{\alpha \beta } = u_0
\begin{bmatrix}
1 & u_0 \\
u_0 & u_0^2 - a_0^2
\end{bmatrix} \\. 
\end{equation}

An acoustic metric and an analogue horizon are based on an
equivalence between Eqs.~(\ref{compact}) and~(\ref{symmat}) on
the one hand and the d'Alembertian equation for a scalar field
in curved geometry on the other (see~\citep{blv11}
and all relevant references therein).
The d'Alembertian equation is
\begin{equation}
\label{alem} 
\Delta \Psi\equiv \frac{1}{\sqrt{-g}}
\partial_\alpha\left({\sqrt{-g}}\, g^{\alpha\beta}
\partial_\beta \Psi\right)=0, 
\end{equation}
where $g^{\alpha \beta}$ is the inverse of the matrix,
$g_{\alpha \beta}$~\citep{vis98,blv11}.
On comparing Eq.~(\ref{compact}) with Eq.~(\ref{alem}), we 
can make the equivalence that 
$h^{\alpha \beta } = \sqrt{-g}\, g^{\alpha \beta}$ and
$g = \det (h^{\alpha \beta})$. Thus, in the non-relativistic
limit the wave equation of $\psi^\prime$ in Eq.~(\ref{psieqnmot}) 
is similar to Eq.~(\ref{alem}). 
The metric that is implicit in Eq.~(\ref{psieqnmot}) is to be
read from Eq.~(\ref{symmat}),
and its inverse establishes an acoustic metric and an
acoustic horizon, when $u_0^2 = a_0^2$~\citep{vis98,blv11}.
In the radial inflow of the non-relativistic~\citet{bon52} accretion, 
this horizon is due to an acoustic black hole.
The radius of the horizon is the critical radius, 
$r_{\mathrm c}=GM/2{a_\mathrm{c}}^2$~\citep{bon52,fkr02},
which cannot be breached by any acoustic wave
propagating against the bulk inflow,
after it originates in the supercritical region,
where $u_0^2 > a_0^2$ and $r < r_{\mathrm c}$.
Therefore, the direction of a wave across the acoustic 
horizon will classically be only inwards~\citep{cgm08}. 
%Outward wave propagation across the 
%horizon is closed classically~\citep{cgm08}. 

The symmetry of Eq.~(\ref{compact}), obtained in the non-relativistic
limit, is lost when we consider the effect of general relativity 
in Eq.~(\ref{psieqnmot}). In this case, 
the curvature of spacetime geometry, whose form is captured by 
the metric coefficient in Eq.~(\ref{statmetric}), is coupled with
the perturbed acoustic field. The coupling gives rise to the source 
term on the right hand side of Eq.~(\ref{psieqnmot}), 
compromising the symmetry of the acoustic metric.
% and violates the acoustic Lorentz invariance. 
Nevertheless, as known for
some fluid analogues of gravity~\citep{rb07,akr20,br21,njar25}, the 
fundamental physics of the acoustic horizon will hold. % qualitatively. 

Some effects of the spacetime geometry on the acoustic field also 
make themselves known through a dispersion relation that can be 
extracted from Eq.~(\ref{psieqnmot}). With respect to a homogeneous
fluid background at rest, %~\citep{vis98}, 
Eq.~(\ref{psieqnmot}) reduces to 
\begin{equation} 
\label{waveq} 
\frac{\partial^2 \psi^\prime}{\partial t^2} 
-f^2 a_0^2 \frac{\partial^2 \psi^\prime}{\partial r^2}
-\frac{a_0^2}{4}\frac{{\mathrm d}f^2}{{\mathrm d}r} 
\frac{\partial \psi^\prime}{\partial r} =0,
\end{equation}
which is the standard form of the wave equation for $f =1$.
A solution, $\psi^\prime (r,t) \sim \exp [i(kr-\omega t)]$, 
tried on Eq.~(\ref{waveq}), gives 
\begin{equation} 
\label{disquad} 
\left(\omega -kv_{\mathrm B}\right)^2 -f^2 a_0^2 k^2
+i \frac{a_0^2}{4} 
\frac{{\mathrm d}f^2}{{\mathrm d}r} k=0, 
\end{equation}
in which $kv_{\mathrm B}$ is due to the bulk motion of the fluid.
The comoving dispersion relation (for $v_{\mathrm B}=0$) is then 
found to be 
\begin{equation} 
\label{disper} 
\omega = \pm fa_0k \left[1-\frac{i}{2k} 
\frac{\mathrm d}{{\mathrm d}r}\left(\ln f\right)\right]^{1/2}.  
\end{equation}
The two roots of Eq.~(\ref{disper}) have the form,
$\omega = \pm fa_0k \left(X+iY\right)$, with $X$ and $Y$ being
real. The imaginary part of $\omega$ contributes to the amplitude 
of the wave. On large radial scales, where curvature is
small, the amplitude is approximately 
\begin{equation}
\label{amplidisp} 
{\big{\vert}} \psi^\prime (r,t) {\big{\vert}} \sim 
\exp \left( \mp \frac{a_0}{4} 
\frac{{\mathrm d}f}{{\mathrm d}r} t \right). 
\end{equation} 
Depending on the sign chosen in Eq.~(\ref{amplidisp}), either 
stability or instability will result. For either outcome the 
spacetime curvature is obviously responsible. However, in flat 
spacetime geometry, where $f =1$, the amplitude will be constant. 

The real part of $\omega$ in Eq.~(\ref{disper}) contributes
to the phase of the wave and and will thus set forth the wave 
velocity. For small spacetime curvature, both the phase 
velocity, $v_{\mathrm p}$, and the group velocity, $v_{\mathrm g}$, 
are approximately   
\begin{equation} 
\label{phasegrav} 
v_{\mathrm p} = \frac{\omega}{k} 
= v_{\mathrm g} = \frac{\partial \omega}{\partial k} 
\simeq \pm fa_0.  
\end{equation}
Clearly, spacetime curvature subdues the speed of the wave. Only 
in flat spacetime ($f=1$) it will be that %far away from the black hole,
$v_{\mathrm p} =v_{\mathrm g} =a_0$. 

%\newpage
\section{Stability under travelling waves}
\label{sec5}
In the non-relativistic (Newtonian) structure of space and time, 
linearized perturbations do not destabilize steady  
accreting solutions~\citep{pso80}. This will be true even with 
general relativistic effects. To see this we design the radial 
perturbation as a high-frequency travelling wave with the solution, 
\begin{equation} 
\label{travsol} 
\psi^\prime (r,t) =\exp \left[i{\mathcal S}(r)-i\omega t \right].
\end{equation} 
The frequency of the wave, $\omega$, is understood to be much greater 
than any characteristic frequency in the system. 
Conversely then, the wavelength, %of the perturbation, 
$\lambda (r)$, % = 2\pi[u_0 \pm a_0(f+u_0^2)^{1/2}]/\omega$,
must have an upper limit, which is given by  
$\lambda (r) = 2\pi[u_0 \pm a_0(f+u_0^2)^{1/2}]/\omega
< r_{\mathrm c}$~\citep{pw2k}. The 
upper limit, $r_{\mathrm c}$, the radius of the acoustic horizon, 
is a natural scale of length in the flow system.  

The solution in Eq.~(\ref{travsol}), inserted into
Eq.~(\ref{psieqnmot}), yields
\begin{multline}
\label{peewkb}
h^{rr}
\left[i\frac{{\mathrm d}^2 {\mathcal S}}{{\mathrm d}r^2}
-\left(\frac{{\mathrm d}{\mathcal S}}{{\mathrm d}r}\right)^2\right]
+\left(i\frac{{\mathrm d}h^{rr}}{{\mathrm d}r}
%\frac{{\mathrm d}{\mathcal S}}{{\mathrm d}r} 
+2 h^{tr} \omega \right) \frac{{\mathrm d}{\mathcal S}}{{\mathrm d}r}
-i\omega \frac{{\mathrm d}h^{rt}}{{\mathrm d}r} -\omega^2 h^{tt} \\ 
= \frac{\mathrm d}{{\mathrm d}r} 
\left[\ln \left(f+u_0^2\right)\right]
\left(-i \omega h^{rt}
+ih^{rr}\frac{{\mathrm d}{\mathcal S}}{{\mathrm d}r}\right). 
\end{multline}
From Eq.~(\ref{peewkb}) we conclude that ${\mathcal S}(r)$ 
has both real and imaginary parts, going as
${\mathcal S}(r)={\mathcal A}(r)+i{\mathcal B}(r)$,
with both $\mathcal A$ and $\mathcal B$ being real. 
Using this form of $\mathcal S$ in Eq.~(\ref{peewkb}), we
separate the real and the imaginary parts, and then
set both equal to zero. The formula 
of $\psi^\prime$ in Eq.~(\ref{travsol}) suggests that  
$\mathcal A$ contributes to the phase of the perturbation and 
$\mathcal B$ to its amplitude. Solutions of both 
$\mathcal A$ and $\mathcal B$ are to be found by a {\it WKB} 
analysis of Eq.~(\ref{peewkb}), which necessitates
${\mathcal A} \gg {\mathcal B}$ and 
$\mathcal B$ to vary much more slowly than $\mathcal A$. 
Hence, we collect the real terms without $\mathcal B$, and 
from the resulting quadratic equation 
in ${\mathrm d}{\mathcal A}/{\mathrm d}r$ we get
\begin{equation} 
\label{kayminus} 
{\mathcal A}=\omega \bigintssss 
\frac{h^{tr} \pm \sqrt{\left(h^{tr}\right)^2-h^{tt}h^{rr}}}
{h^{rr}}\,{\mathrm d}r.
\end{equation}
Likewise, from the imaginary part, on using Eq.~(\ref{kayminus}), 
we get
\begin{equation} 
\label{kaynot} 
{\mathcal B} = \ln \left[
\frac{\sqrt{\left(h^{tr}\right)^2-h^{tt}h^{rr}}}{f+u_0^2}   
\right]^{1/2}.
\end{equation} 

Now we carry out a self-consistency check that 
${\mathcal A} \gg {\mathcal B}$, as a basic requirement of the 
{\it WKB} analysis. First we note that 
$\mathcal A$ contains $\omega$ (the high frequency of the travelling
wave), and in this respect is of a
leading order over $\mathcal B$, which contains $\omega^0$. Next,
on large radial scales, where $f \longrightarrow 1$,
the asymptotic behaviour of the background velocity is
$u_0 \sim r^{-2}$ and the speed of sound, $a_0$, becomes 
constant~\citep{fkr02}. 
Under these conditions, the asymptotic values of $h^{\alpha \beta}$ 
show that ${\mathcal A} \sim \omega r$
and ${\mathcal B} \sim \ln r$. Considering all of these facts together, 
we see that our solution scheme conforms to the 
{\it WKB} prescription.

Using Eqs.~(\ref{metelem}) for all $h^{\alpha \beta}$ in 
Eqs.~(\ref{kayminus})~and~(\ref{kaynot}), the solution 
in Eq.~(\ref{travsol}), %will be %is %expanded to be     
with $u_0 = -\vert u_0 \vert$ for the inflow, is expanded to be   
\begin{multline} 
\label{psisol} 
\psi^\prime (r,t) \sim \frac{\left(f+u_0^2\right)^{1/4}}
{\sqrt{a_0 \vert u_0 \vert}} \\
\times \exp\left[i\omega \bigintssss %\int
\frac{u_0 f^{-1}\sqrt{f+u_0^2}\left(1-a_0^2\right)\pm a_0}
{u_0^2- a_0^2\left(f+u_0^2\right)}  
%\left( 
%\frac{1}{\sqrt{f+u_0^2} \left(u_0 \mp \sqrt{f+u_0^2}a_0\right)} 
%+\frac{u_0}{f\sqrt{f+u_0^2}}
%\frac{h^{tr} \pm \sqrt{\left(h^{tr}\right)^2-h^{tt}h^{rr}}}{h^{rr}}
%\right)
\,{\mathrm d}r \right]e^{-i\omega t}.  
\end{multline} 
In the phase part of Eq.~(\ref{psisol}) the positive and negative signs, 
placed together, 
indicate a superposition of incoming waves (for the negative sign) and 
outgoing waves (for the positive sign), respectively. 
The amplitude of the wave is subdued because of the coupling between
the spacetime geometry and the flow. In the Newtonian limit, 
the amplitude 
is $\vert \psi^\prime \vert \sim (a_0 \vert u_0 \vert)^{-1/2}$~\citep{pso80}. 
Either way, the amplitude does not diverge and the steady background 
flow is not destabilized. % by the perturbation. 

\section{Tunnelling and Hawking radiation} 
\label{sec6}
For steady radial inflows $u_0 =-\vert u_0 \vert$. Waves travelling 
outwards against such inflows encounter a singularity when 
$\vert u_0 \vert =a_0 (f+u_0^2)^{1/2}$, as should be obvious from the 
integrand in the phase part of Eq.~(\ref{psisol}). 
Circumvention of the singularity requires
rendering it as a simple pole on the path of the integration,
and then applying Cauchy's residue theorem on the path.
To carry out this procedure, we recast Eq.~(\ref{kayminus}) as
\begin{equation} 
\label{phasesingu} 
{\mathcal A}= \omega \bigintssss \left[
\frac{1}{\sqrt{f+u_0^2}
\left(u_0 \mp a_0 \sqrt{f+u_0^2}\right)} %\,{\mathrm d}r.
+ %\omega \bigintssss 
\frac{u_0}{f\sqrt{f+u_0^2}} \right]\,{\mathrm d}r.
\end{equation}
We read Eq.~(\ref{phasesingu}), with the two terms on its
right hand side, as 
${\mathcal A} = {\mathcal A}_1 + {\mathcal A}_2$. 
The singularity is contained in the first term, ${\mathcal A}_1$. 
Hence, we now consider only this term with its lower sign, since 
this sign pertains to a wave propagating outwards against the 
radial inflow. Accordingly, we write 
\begin{equation} 
\label{singu1} 
{\mathcal A}_1 = \omega \bigintssss 
\frac{{\mathrm d}r}
{\sqrt{f+u_0^2}
\left(a_0 \sqrt{f+u_0^2}-\vert u_0 \vert \right)}. 
\end{equation}
The main contribution to the integral in Eq.~(\ref{singu1}) comes 
from the immediate neighbourhood of the singularity, where 	
$r=r_{\mathrm c}$ and $\vert u_0 \vert =a_0 (f+u_0^2)^{1/2}$, 
as shown in Eq.~(\ref{critcon1}). 
A Taylor expansion about the singularity, up to the first order, 
goes as
\begin{multline} 
\label{taylor} 
a_0\sqrt{f+u_0^2} -\vert u_0 \vert \simeq 
\left[
a_{\mathrm c}\sqrt{f(r_{\mathrm c})+u_{\mathrm c}^2}
-\vert u_{\mathrm c} \vert \right] 
\\
+\left[\frac{\mathrm d}{{\mathrm d}r}
\left(a_0\sqrt{f+u_0^2}-\vert u_0 \vert \right)\right]_{r_{\mathrm c}}
(r-r_{\mathrm c}). % + \ldots 
\end{multline} 
Going by Eq.~(\ref{critcon1}), the zero-order term in the Taylor
expansion vanishes. Consequently, we approximate Eq.~(\ref{singu1}) as
\begin{equation}
\label{alphahor} 
{\mathcal A}_1 \simeq 
\frac{\omega a_{\mathrm c}}
{a_{\mathrm c}\sqrt{f(r_{\mathrm c})+u_{\mathrm c}^2} 
\left[{\mathrm d}\left(a_0\sqrt{f+u_0^2}-\vert u_0 \vert \right){\big{/}}
{\mathrm d}r \right]_{r_{\mathrm c}}}
\int \frac{{\mathrm d}r}{r-r_{\mathrm c}}.
\end{equation}
In effect, the Taylor expansion has transformed the singularity 
at $\vert u_0 \vert =a_0 (f+u_0^2)^{1/2}$
to a simple pole at $r=r_{\mathrm c}$. Now we recall 
%from Sec.~\ref{sec4} following Eq.~(\ref{alem}), 
that the singularity defines the horizon of an acoustic black hole
in the radially accreting fluid.
At the horizon, the analogue surface gravity is %~\citep{vis98} 
\begin{equation} 
\label{surfgrav} 
G_{\mathrm s}= a_{\mathrm c}
\sqrt{f(r_{\mathrm c})+u_{\mathrm c}^2}
\left[\frac{\mathrm d}{{\mathrm d}r}
\left(a_0\sqrt{f+u_0^2}-\vert u_0 \vert \right)\right]_{r_{\mathrm c}}
\end{equation} 
and the %analogue 
Hawking temperature, enhanced by curvature, is %~\citep{vis98}  
%\begin{multline} 
\begin{equation} 
\label{hawktemp} 
T_\mathrm{H}=\frac{\hbar G_\mathrm{s}}{2\pi k_\mathrm{B} 
a_{\mathrm c}\sqrt{f(r_{\mathrm c})+u_{\mathrm c}^2}}
=\frac{\hbar G_\mathrm{s}\sqrt{1+3a_{\mathrm c}^2}}{2\pi k_\mathrm{B} 
a_{\mathrm c}}. % \\
%= \frac{\hbar}{2\pi k_\mathrm{B}} 
%\left[\frac{\mathrm d}{{\mathrm d}r}
%\left(a_0\sqrt{f+u_0^2}-\vert u_0 \vert \right)\right]_{r_{\mathrm c}}
%\end{multline} 
\end{equation} 
In terms of $G_\mathrm{s}$ and
$T_\mathrm{H}$, the integral in Eq.~(\ref{alphahor}),
on extracting the residue at the pole~\citep{dk96}, is reduced to  
\begin{equation} 
\label{alphres0} 
{\mathcal A}_1 \simeq 
\frac{\omega 
a_{\mathrm c}}{G_\mathrm{s}} 
\left(\pm i \pi \right)+{\mathcal P}\left[{\mathcal A}_1 \right] 
=\frac{\hbar \Omega}{2k_\mathrm{B}T_\mathrm{H}
%\sqrt{f(r_{\mathrm c})+u_0^2(r_{\mathrm c})}
} 
\left(\pm i\right)+{\mathcal P}\left[{\mathcal A}_1 \right], 
\end{equation}
in which ${\mathcal P}[{\mathcal A}_1]$ is the
principal value of the integral and 
\begin{equation}
\label{tunfreq} 
\Omega 
= \frac{\omega} {\sqrt{f(r_{\mathrm c})+u_{\mathrm c}^2}} 
= \omega \sqrt{1+3a_{\mathrm c}^2}.
\end{equation}

The imaginary part of
${\mathcal A}_1$ in Eq.~(\ref{alphres0}) contributes to the
amplitude. The negative sign in
$\pm i$ is due to a clockwise detour of
the pole, and the positive sign is due to an anti-clockwise
detour. Both signs are mathematically valid, 
and the choice of a sign %appropriate sign 
depends on the physical boundary condition at the pole~\citep{dk96}.
%Noting the role of $\mathcal A$, through $\mathcal S$, in 
%Eq.~(\ref{travsol}), we realize that 
%The imaginary part of 
%${\mathcal A}_1$ in Eq.~(\ref{alphres0}) contributes to the 
%amplitude.  % of $\psi$. 
Now, outgoing acoustic waves will be blocked just inside the 
sonic horizon because here the wavenumber,
$k(r)= %2\pi/\lambda(r) =
\omega/[\vert  u_0 \vert -a_0(f+u_0^2)^{1/2}]$, undergoes 
an infinite blue-shift. 
%the horizon of the acoustic black hole 
%acts like an unyielding barrier to outgoing acoustic waves. 
This 
boundary condition at the horizon necessitates the choice 
of the positive sign of $i$ in Eq.~(\ref{alphres0}), and as such, 
the wave, which must be very weak
with a decaying amplitude, tunnels through the barrier.
The tunnelling amplitude is 
\begin{equation} 
\label{amplipen} 
\big{\vert} \psi_\mathrm{T}^\prime \big{\vert}
\sim \exp \left( - 
\frac{\hbar \Omega}{2k_\mathrm{B}T_\mathrm{H}}\right)
\end{equation} 
and the tunnelling probability is
$\vert \psi_\mathrm{T}^\prime \vert^2$. 
In the amplitude, $\hbar \Omega$
is scaled by the analogue Hawking temperature, 
$T_\mathrm{H}$, which makes the tunnelling process a case 
of acoustic Hawking radiation. %~\citep{vis98,blv11}.
The frequency of the Hawking phonons, $\Omega$, is 
enhanced by the spacetime geometry, as Eq.~(\ref{tunfreq}) shows. 

\section{Remarks}
\label{sec7}
%The wavenumber, 
%$k(r)= %2\pi/\lambda(r) =
%\omega/[\vert  u_0 \vert -a_0(f+u_0^2)^{1/2}]$, indicates an 
%arbitrary blue-shift for waves near the acoustic horizon.  
Waves against the inflow are arbitrarily blue-shifted near 
the acoustic horizon. 
However, in a non-ideal fluid, a physical cut-off to the blue-shift
is imposed by viscosity~\citep{akr20} 
and surface tension~\citep{njar25}. 
In~\citet{bon52} 
accretion with a small viscous correction, viscosity creates
a thin layer of uncertainty about the sonic horizon and 
allows an outgoing wave packet to avoid an infinite 
blue-shift~\citep{akr20}. Surface 
tension works similarly for waves at the horizon of a 
hydrodynamic white hole along a circular hydraulic jump~\citep{njar25}.  
In our study of an ideal fluid here, we see that although 
the coupling between the spacetime geometry and the flow acts like 
a non-ideal effect on the acoustic metric, it does not prevent
an arbitrary blue-shift.

In passing we point out that steady transonic accretion 
lies on the separatrix of a dynamical system and is extremely 
sensitive to boundary conditions~\citep{rb02,rbcqg07}. Transonic 
flows can be physically realized only through
dynamics~\citep{rb02,rbcqg07,rr07,sr14}. 

\begin{acknowledgments}
This research has made use of NASA's Astrophysics Data System.
The authors thank J. K. Bhattacharjee and T. Naskar 
for useful comments.  
\end{acknowledgments}

\bibliography{jrHM2026_rev} 
\end{document}